# Partitioned Hankel-based Diffusion Models for Few-shot Low-dose CT Reconstruction

Wenhao Zhang, Bin Huang, Shuyue Chen, Xiaoling Xu, Weiwen Wu, *Member, IEEE*
Qiegen Liu, *Senior Member, IEEE*

***Abstract*—Low-dose computed tomography (LDCT) plays a vital role in clinical applications by mitigating radiation risks. Nevertheless, reducing radiation doses significantly degrades image quality. Concurrently, common deep learning methods demand extensive data, posing concerns about privacy, cost, and time constraints. Consequently, we propose a few-shot low-dose CT reconstruction method using Partitioned Hankel-based Diffusion (PHD) models. During the prior learning stage, the projection data is first transformed into multiple partitioned Hankel matrices. Structured tensors are then extracted from these matrices to facilitate prior learning through multiple diffusion models. In the iterative reconstruction stage, an iterative stochastic differential equation solver is employed along with data consistency constraints to update the acquired projection data. Furthermore, penalized weighted least-squares and total variation techniques are introduced to enhance the resulting image quality. The results approximate those of normal-dose counterparts, validating PHD model as an effective and practical model for reducing artifacts and noise while preserving image quality.**

## I. Introduction

Computed tomography (CT) techniques have transformed clinical settings, offering valuable contributions to diagnosis and intervention procedures, including imaging, image-guided needle biopsy, image-guided intervention, and radiotherapy [1]. However, the widespread usage of CT scans has raised concerns regarding the potential risk of cancer associated with X-ray radiation dose. To mitigate this concern, the method of adjusting the tube current is employed to reduce X-ray exposure and consequently lower the radiation dose received by patients during CT examinations [2, 3].

The various proposed algorithms for low-dose CT image reconstruction can be classified into three types: image post-processing methods [4, 5], sinogram domain methods [6-8], and iterative reconstruction methods [9, 10]. Image post-processing methods enable the direct enhancement of low-quality images without utilizing raw projection data. Chen *et al.* [4] combined the autoencoder, deconvolution network, and shortcut connections into the residual encoder-decoder convolutional neural network (RED-CNN) to reconstruct CT images with structural preservation and noise suppression. Ding *et al.* [5] proposed a low-dose CT image reconstruction method based on deep learning regularization. This method unfolds a proximal forward-backward splitting (PFBS) framework with data-driven image regularization via deep neural networks. However, removing severe streak artifacts and accurately recovering image details and features without projection data can be challenging. The sinogram domain-based reconstruction methods are helpful in solving this problem. Yin *et al.* [6] introduced a domain progressive 3D residual convolution network (DP-ResNet) for the LDCT imaging procedure, while Balda *et al.* [7] combined the specific properties of the CT measurement process to design a structure adaptive sinogram (SAS) filter. Manduca *et al.* [8] proposed a bilateral filtering method incorporating the CT noise model, which achieved higher noise resolution compared to commercial reconstruction kernels. Iterative reconstruction focuses on solving the low-dose CT problem iteratively by extracting prior information from target images [9]. Various priors were developed, with total variation (TV)-based reconstructions [10-12] being the most well-known. Tian *et al.* [10] developed an iterative CT reconstruction algorithm with edge-preserving TV regularization, which prioritized smoothing in the non-edge regions of the image. Additionally, Zhang *et al.* [11] proposed a model based on fractional-order TV, which replaced traditional TV, thus suppressing the over-smoothing effect. Sagheer *et al.* [12] considered tensor TV and developed a method based on low-rank approximation to improve global smoothness. However, iterative reconstruction methods seriously affect computational costs, striking a balance between image quality and computational efficiency in practical applications.

Diffusion models have provided a fresh perspective for image processing tasks [13, 14]. Li *et al.* [15] introduced an unsupervised image domain score generation model for low-dose CT reconstruction. Song *et al.* [16] incorporated forward and backward diffusion processes into a framework based on stochastic differential equations (SDE) to design a score-based generative model. This model has found extensive application in clinical medical scenarios and has demonstrated remarkable performance [17, 18]. Moreover, diffusion models exhibit higher efficiency when compared to generative adversarial networks (GANs).

Collecting medical data is a complex and expensive task, and the ongoing issues related to privacy and security add to the complexity of sharing clinical data. Consequently, there is a shortage of readily available clinical data, prompting researchers to explore few-shot reconstruction studies such as image denoising [19], image classification [20], and image segmentation [21]. To overcome this challenge, chang-

---

This work was supported in part by National Natural Science Foundation of China under 62122033. (W. Zhang and B. Huang are co-first authors) (Corresponding authors: W. Wu and Q. Liu)

W. Zhang, X. Xu and Q. Liu are with School of Information Engineering, Nanchang University, Nanchang 330031, China. (zhangwenhao@email.ncu.edu.cn, {xuxiaoling, liuqiegen}@ncu.edu.cn)

B. Huang, S. Chen are with School of Mathematics and Computer Sciences, Nanchang University, Nanchang 330031, China. ({huangbin, shuyue.chen}@email.ncu.edu.cn)

W. Wu is with School of Biomedical Engineering, Sun Yat-Sen University, Shenzhen, Guangdong, China. (wuweiw7@mail.sysu.edu.cn)

ing the scale of data has become a widely adopted solution. These methods often involve downsizing and sampling larger images to capture a variety of underlying statistical properties, which are subsequently employed in training datasets. Due to its data redundancy and structural reproducibility, Hankel matrix is commonly employed in various fields such as image denoising [22], artifact removal [23], and compressed sensing [24]. Wang *et al*. [25] designed an encoder-decoder network based on Hankel matrix decomposition, which leverages low-rank Hankel matrices for reconstructing high-dose images from low-dose images using few-shot learning techniques. However, despite research on few-shot natural images is emerging, studies related to medical images are still relatively limited [26, 27].

In this study, we introduce a few-shot learning technology called Partitioned Hankel-based Diffusion (PHD) for low-dose CT reconstruction using projection domain diffusion modeling. The method combines the low-rank structural-Hankel matrix with the diffusion model to generate an ideal sinogram from low-dose projection data. Additionally, we employ penalized weighted least-squares (PWLS) and total variation regularization to enhance image quality and accelerate iteration speed. Unlike previous supervised learning methods, this work proposed unsupervised approach does not require retraining of low-dose/normal-dose CT image pairs when there are changes in projection dosage. In addition, training only requires a few-shot data samples. Constructing a Hankel matrix from projected domain data helps mitigate the significant differences between CT images in the image domain, thereby aiding in better generalization during model training to improve model performance.

The theoretical and practical contributions of this work can be summarized as follows:

- We propose a few-shot low-dose CT reconstruction method in the projection domain, which utilizes partitioned Hankel-based diffusion models to learn the data distribution of each Hankel matrix separately. Compared to a single Hankel matrix, parallel computing is employed to reduce computational complexity. Additionally, the multiply models learn different parts of the data, thereby enhancing the representation capacity of prior information.
- The diffusion model is used in projection domain. Specifically, leveraging the data structure redundancy in the Hankel domain to transform individual data into multi-center data. Each group of data is subsequently subjected to diffusion modeling for further distribution learning. Sufficient prior information can be extracted with only a small number of samples.
- The input of the PHD model shifts from pure noise to low-dose CT projection data, significantly reducing the number of iterations with the assistance of the PWLS fidelity module and the TV denoising module. With just 10 steps, we can achieve perfect reconstruction of fine details, and even with only 1 step, the results are remarkably impressive. This improvement substantially cuts down on the reconstruction time.

The remaining sections of this study are organized as follows: Section II provides background information on score-based diffusion models and the construction process of the Hankel matrix. In Section III. The experimental results and analysis are shown in Section IV. The discussion is given in Section V. Finally, we draw conclusions in Section VI.

## II. Preliminary

### A. Low-dose CT Imaging

Low-dose CT reconstruction attempts to reconstruct the blurred parts in low-dose CT data, which is a classic inverse problem. Currently, there are many algorithms proposed for low-dose CT reconstruction, which can be roughly divided into three categories: image post-processing methods (which can directly skip the processing of raw projection data and process the low-quality image), sinogram domain methods, and iterative reconstruction methods. Specifically, assuming that $x \in \mathbb{R}^M$ is the degraded sinogram, the forward formula of the sinogram reconstruction problem can be given by the following equation:

$$y = x + n \tag{1}$$

where $n \in \mathbb{R}^M$ represents additive noise and $y$ denotes a low-dose sinogram. It should be noted that the inverse problem refers to solving $x$ from $y$ .

To avoid ill-posed issues, the CT sinogram reconstruction problem is represented by an optimization equation with the following constraint:

$$\min_x \{\| x - y \|_2^2 + \mu R(x)\} \tag{2}$$

where $\| x - y \|_2^2$ is the data fidelity term. $R(x)$ represents the adjustment prior degree term, which is selected as the TV semi-norm. $\|\bullet\|_2^2$ represents the $\ell_2$-norm. Besides, $\mu$ is a factor to keep a good balance between the data-consistency (DC) term and regularization term.

Considering the TV aspect, this optimization process distinguishes the infinite solutions of (2) and selects the best image with desired properties as the reconstructed sinogram. Typically, the TV term is defined as:

$$R(x) = \|x\|_{TV}^2 = \int_\Omega |\nabla x| dx \tag{3}$$

where $\Omega$ is the bounded domain. $\nabla x$ represents the gradient of the sinogram $x$ . The TV term is robust in removing noise and artifacts.

### B. Construction of Hankel Matrix

Due to the ability of Hankel matrices to fully exploit the redundancy of identical pixels between different positions, Peng *et al*. [28] constructed a Hankel matrix from the sample data to alleviate the problem of few-shot data samples. Since projection data can be represented by low-rank Hankel matrix, the internal relationship between the sinogram and the Hankel matrix is illustrated in Fig. 1.

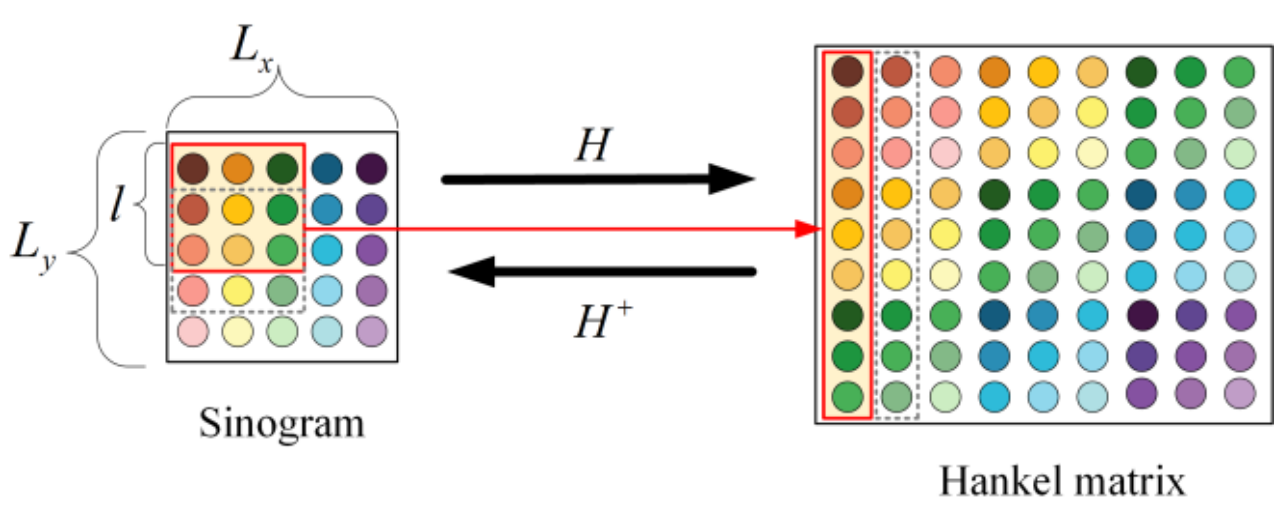

**Fig. 1.** Constructing a new Hankel matrix formulation from a sinogram through $H$ and vice versa $H^+$ .

As shown in Fig. 1, the sinogram $L_x \times L_y$ undergoes transformation $H(\bullet)$ , resulting in the construction of the Hankel matrix of size $l^2 \times (L_x - l + 1)(L_y - l + 1)$ . Fig. 1 demonstrates the case of $L_x = 5$, $L_y = 5$, and $l = 3$. Additionally, the step size of the sliding window is set to 1, and the window size is set to $l \times l$.

In the new Hankel matrix formulation, individual block data in the projection domain is vectorized as columns. The linear operator $H(\bullet)$ is defined as the operation to construct the Hankel matrix from the sinogram:

$$H = R^{L_x \times L_y} \to R^{l^2 \times (L_x - l + 1)(L_y - l + 1)} \tag{4}$$

When performing the reverse operation $H^{+}(\bullet)$ from (4) to generate the sinogram, multiple anti-diagonal entries are averaged and stored in the projection domain, represented as follows:

$$H^{+} = R^{l^2 \times (L_x - l + 1)(L_y - l + 1)} \to R^{L_x \times L_y} \tag{5}$$

where the superscript $+$ represents the pseudo-inverse operator. It is equivalent to averaging the anti-diagonal elements and placing them in the appropriate locations.

### C. Score-based SDE

The impressive success of diffusion models, specifically score-based stochastic differential equations (SDE), in producing realistic and varied image samples has sparked widespread interest [16]. Score-based SDE encompasses both the forward process and the reverse-time process.

Consider a continuous diffusion process $\{x(t)\}_{t=0}^{T}$ with $x(t) \in \mathbb{R}^N$, where $t \in [0, T]$ represents the progression time variable. $N$ denotes the dimension of the sinogram. The forward diffusion process can be expressed as the solution to the following SDE:

$$dx = f(x,t)dt + g(t)dw \tag{6}$$

where $f(x,t) \in \mathbb{R}^N$ and $g(t) \in \mathbb{R}$ correspond to the drift coefficient and diffusion coefficient, respectively. $w \in \mathbb{R}^N$ induces the Brownian motion.

Via reversing the above process, samples can be attained. Notably, the reverse-time SDE is also a diffusion process, which could be expressed as follows:

$$dx = [f(x,t) - g(t)^2 \nabla_x \log p_t(x)]dt + g(t)d\bar{w} \tag{7}$$

where $dt$ is the infinitesimal negative time step, $\bar{w}$ is a standard Wiener process when time flows backward from $T$ to 0, and $\nabla_x \log p_t(x)$ is the score of each marginal distribution.

## III. Proposed Method

### A. Low-dose CT Imaging Model

As shown in Fig. 2, the linear measurement process for low-dose CT imaging is quite intuitive. Intuitively, $n$ can be defined as low-dose noise on a sinusoidal curve. $I$ represents CT image.

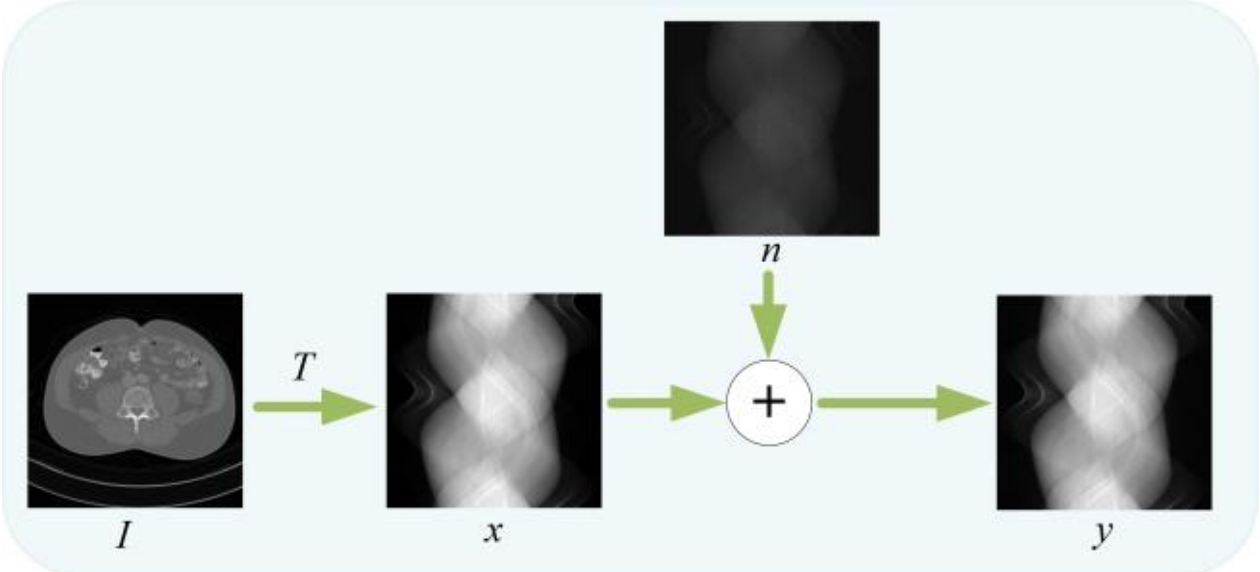

**Fig. 2.** Linear measurement process for low-dose CT.

If the appropriate $x$ of size $768 \times 768$ is measured in the presence of $n$, the CT imaging reconstruction problem can be solved by the following equation:

$$y = T(I) + n = x + n \tag{8}$$

where $T(\bullet)$ corresponds to the Radon transform, $y$ is the low-dose CT sinogram with size $768 \times 768$.

Based on extensive experimental support with a significant amount of real projection data, low-dose sine images can be approximated as ideal projection data corrupted by additional noise [29]. It is assumed that only a monochromatic light source is used, and the additional noise follows a Poisson distribution. Specifically, the Poisson model for intensity measurements is expressed as follows:

$$L_i \sim Poisson\{a_i e^{-[x]_i} + r_i\}, \quad i = 1, \cdots, N_m \tag{9}$$

where $L_i$ represents the transmitted number of photons, $a_i$ is the X-ray source intensity of the $i$ -th ray, and $r_i$ signifies the background contribution from scatter and electronic noise. $x$ corresponds to the vector representing the attenuation coefficients with inverse length units, $N_m$ stands for the number of measurements. In (9), the noise level is characterized by $a_i$, its specific value is provided in Section IV.

The measurement data is transformed into a weighted Gaussian form through a logarithmic operation:

$$y_i \sim N([x]_i, \bar{L}_i / (\bar{L}_i - r_i)^2) \tag{10}$$

where $\bar{L}_i = E[L_i]$. In fact, the low-dose CT problem is formulated as a typical ill-posed inverse problem. To address this issue, the posterior distribution $p(x \mid y)$ is introduced by the theory of Bayesian inversion [30]. Consequently, the inverse problem is transformed into a measurement $y$.

### B. Training Stage

During the training phase, the Hankel matrix is preprocess and input into the network, as shown in Fig. 3. In order to store information within the data, a window of size $8 \times 8$ is slid over the initial data, resulting in the construction of a Hankel matrix of size $579121 \times 64$. After applying the Hankel transform, the same information appears at different positions within the matrix, harnessing its redundancy to capture internal statistical information. It is important to note that during the training process, one or few normal-dose sinograms are required. The Hankel matrix is constructed from the projection data as follows:

$$H_s = H(x) \tag{11}$$

where the initial dose projection data is represented as $x$, $H(\bullet)$ corresponds to the Hankel transform, and $H_s$ represents the constructed Hankel matrix.

During the training stage, three local Hankel matrices are extracted from the original Hankel matrix. The first Hankel matrix, with a size of $289560 \times 64$, retains the left half of the information from the original Hankel matrix. The second Hankel matrix, with a size of $289561 \times 64$, captures the right half of the information. Lastly, the size of the third Hankel matrix is $289560 \times 64$, preserving the middle portion of the information from the original Hankel matrix.

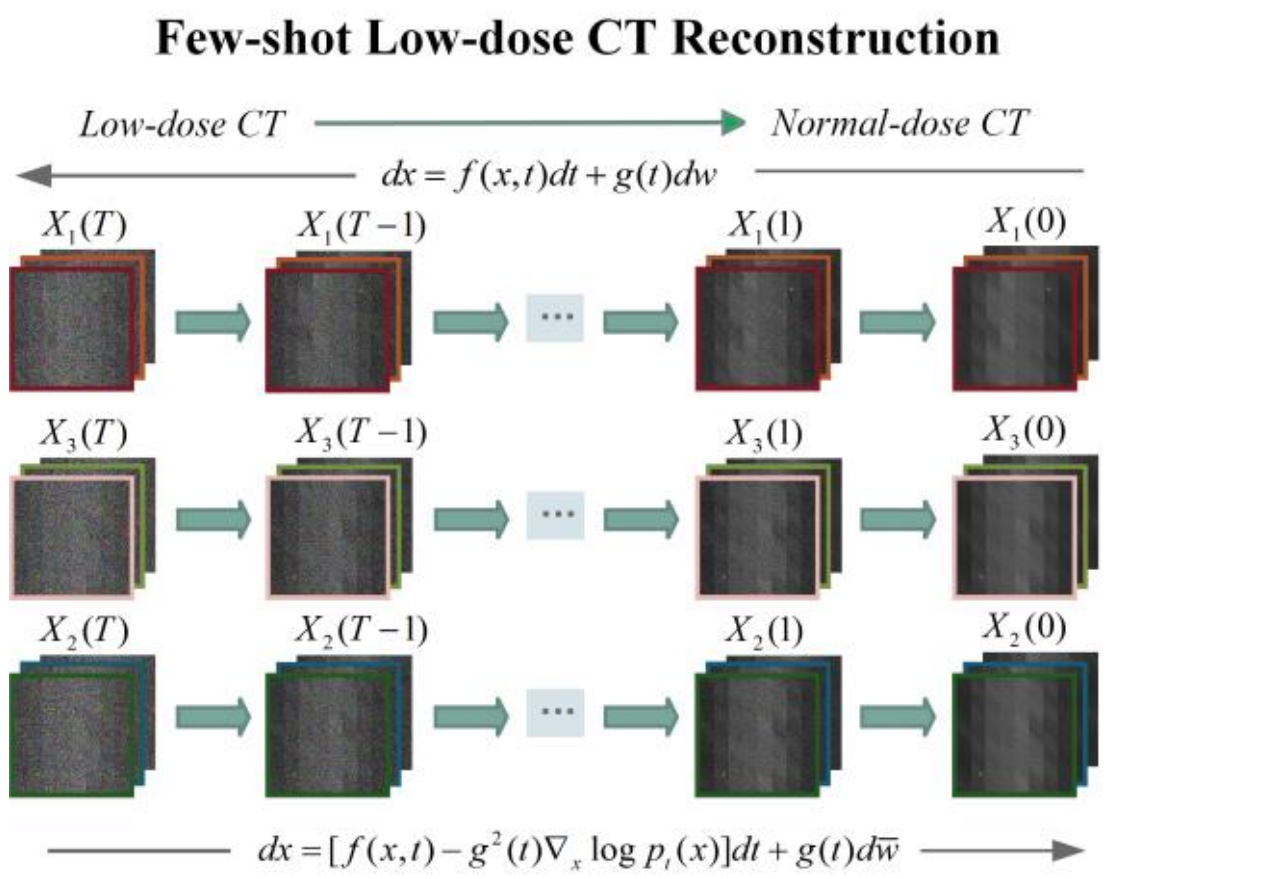

**Fig. 3.** The perturbed data by noise is smoothed according to the trajectory of an SDE. By estimating the score function $\nabla_x \log p_t(x)$ with SDE, it is possible to approximate the reverse SDE and then solve it to generate sinogram samples from noise.

These three Hankel matrices are processed in parallel and cropped into high-dimensional data of size $64\times64\times4524$. In addition, multiple small patches are randomly extracted from the matrix as illustrated in Fig. 4. Specifically, the constructed Hankel matrix is partitioned into small blocks using the following process:

$$X = S(H_s) \tag{12}$$

This random partition operation, denoted by the symbol $S(\bullet)$ is applied to the high-dimensional tensor $X$ to extract numerous small blocks. Then these blocks are considered as inputs to the network, resulting in the construction of a large number of tensors. This data augmentation technique enhances the training set, enabling the network to acquire a sufficient amount of prior knowledge.

The score-based model described in Section II. C learns a prior distribution by leveraging an SDE. This is achieved through a forward SDE, which gradually injects noise, smoothly transforming complex data distributions into the known prior distribution. Fig. 3 illustrates these two processes.

During the training phase, the parameter $\theta^*$ of the scoring network is optimized to achieve the peak performance of the network. The objective function can be described as follows:

$$\begin{aligned}\theta^* = \arg\min_{\theta} \mathbb{E}_t\{\lambda(t)\mathbb{E}_{x(0)}\mathbb{E}_{x(t)|x(0)}[ \\ \left\| s_\theta(x(t),t) - \nabla_{x(t)} \log p_t(x(t) \mid x(0)) \right\|_2^2 ]\}\end{aligned} \tag{13}$$

where $\lambda:[0,T]\to\mathbb{R}^+$ is a positive weighting function and $t$ is uniformly sampled over $[0,T]$. $p_t(x(t)\mid x(0))$ is the Gaussian perturbation kernel centered at $x(0)$. Once the network satisfies $s_\theta(x(t),t) \simeq \nabla_x \log p_t(x)$, $\nabla_x \log p_t(x)$ will be known for all $t$ by solving $s_\theta(x(t),t)$.

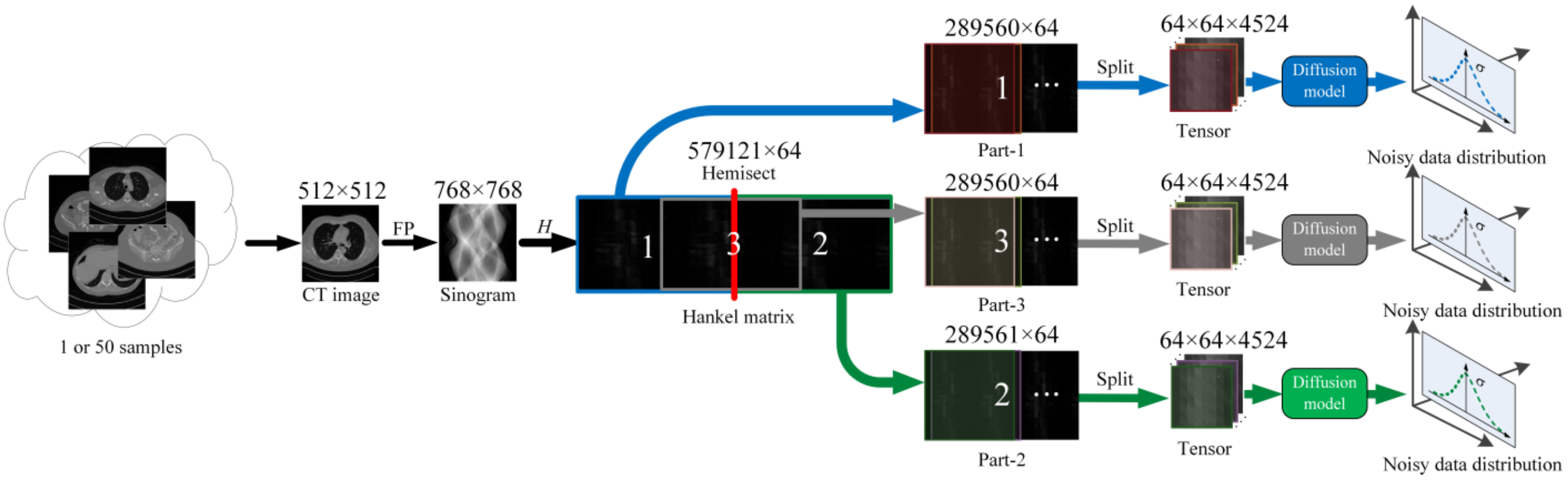

**Fig. 4.** The training stage of PHD. First, the projection data is used to construct a low-rank Hankel matrix. Next, three subsets of Hankel matrices are extracted from the Hankel matrix for parallel processing. Finally, each of the three subsets of Hankel matrices is randomly divided into multiple tensors, and denoising score matching is applied to train the network to learn the gradient distribution.

### *C. Iterative Reconstruction Stage*

This part describes the iterative reconstruction process of the PHD model. First, through the $T^{-1}(\bullet)$ operation (*i.e.*, the inverse Radon transform), sinogram data is obtained from the CT image $I$, corresponding to the forward projection (FP):

$$x = T^{-1}(I) \tag{14}$$

The score-based diffusion model is employed to estimate the prior distribution of sinogram data $p_t(x)$, enabling noise suppression and enhanced information. Unlike perturbing data with a finite number of noise distributions, this method takes into account continuous distributions over time during the forward diffusion process. By reversing the SDE, random noise can be transformed into data suitable for sampling. In this paper, as suggested in [31], Predictor-Corrector (PC) sampling is introduced in the sample update step. In PC sampling, the predictor is treated as a numerical solver for the reverse-time SDE. Once the reverse-time SDE process concludes, samples are generated according to the discretized prior distribution, which can be discretized as follows:

$$\begin{aligned} x^i \leftarrow x^{i+1} + (\sigma_{i+1}^2 - \sigma_i^2) s_\theta(x^{i+1}, \sigma_{i+1}) + \sqrt{\sigma_{i+1}^2 - \sigma_i^2}\, z \\ i = N-1, \cdots, 0 \end{aligned} \tag{15}$$

where $z \sim \mathcal{N}(0,1)$, $x(0) \sim p_0$, and $\sigma_0 = 0$ are chosen to simplify the notation. The above formulation is repeated for $i = N-1,\cdots,0$. Thus, previous discrete process turns into continuous diffusion process. With adding the conditional constraints to (15), it can be rewritten as follows:

$$\begin{aligned} x^i = x^{i+1} + (\sigma_{i+1}^2 - \sigma_i^2)\nabla_x[\log p_t(y\mid x^{i+1}) + \log p_t(x_{LR}^{i+1}) \\ + \log p_t(x_{TV}^{i+1})] + \sqrt{\sigma_{i+1}^2 - \sigma_i^2}\, z \end{aligned} \tag{16}$$

In the above equation, $\log p_t(y\mid x)$ stems from sinogram data knowledge. $\log p_t(x_{LR})$ is derived from low-rank (LR) prior and $\log p_t(x_{TV})$ comes from TV prior.

Next, following the Hankel transform (HT) operation, each updated sample $x$ is further transformed into three distinct new Hankel matrices $H_1^i$, $H_2^i$, and $H_3^i$.

$$[H_1^i, H_2^i] = H_p^i \leftarrow H(x^i) \tag{17}$$

The local subset of $H_s^i$ is described as follows:

$$\begin{aligned} H_{sL}^i &= H_s^i(0:[m/2], 64) \\ H_{sR}^i &= H_s^i([m/2]:m, 64), s = 1,2,3. \end{aligned} \tag{18}$$

where $H_{sL}^i$ and $H_{sR}^i$ represent the left/right half of $H_s^i$. The size of the Hankel matrix $H_s^i$ is defined as $m \times 64$. Specifically, when $s=1$, the value of m is 289560. When $s=2$, the value of m is 289561. $H_3^i$ is obtained by concatenating $H_{1R}^i$ and $H_{2L}^i$ as follows:

$$H_3^i = [H_{1R}^i, H_{2L}^i] \tag{19}$$

where $H_3^i$ is a Hankel matrix of size $289560 \times 64$.

***LR Step:*** To facilitate processing and analysis, the first step is to decompose the Hankel matrices $H_1^i$ and $H_2^i$ using Singular Value Decomposition (SVD). Simultaneously calculate the average of $H_{1R}^i$ and $H_{3L}^i$, and represent the results as $H_{2L}^i$ and $H_{3R}^i$, respectively.

$$\begin{aligned} \bar{H}_{1R}^i &= [svd(H_{1R}^i) \oplus svd(H_{3L}^i)]/2 \\ \bar{H}_{2L}^i &= [svd(H_{2L}^i) \oplus svd(H_{3R}^i)]/2 \end{aligned} \tag{20}$$

where $\oplus$ represents adding the corresponding elements in two matrices. $\bar{H}_{1R}^i$ and $\bar{H}_{2L}^i$ are matrices of size $144780 \times 64$. The matrix elements of $\bar{H}_{1R}^i$ are obtained by taking the average of corresponding elements from $H_{1R}^i$ and $H_{3L}^i$. Similarly, the matrix elements of $\bar{H}_{2L}^i$ are obtained by taking the average of corresponding elements from $H_{2L}^i$ and $H_{3R}^i$.

$$[U \Delta V^T] = svd[H_{1L}^i, \bar{H}_{1R}^i, \bar{H}_{2L}^i, H_{2L}^i]^T \tag{21}$$

where $U$ is an orthogonal matrix, $\Delta$ is a diagonal matrix with non-negative diagonal elements, and $V$ is an orthogonal matrix. Specifically, $H_1^i$ is a $289560 \times 64$ Hankel matrix and $H_2^i$ is a $289561 \times 64$ Hankel matrix of rank $L$ while $U_{[k]}$, $\Delta_{[k]}$ and $V_{[k]}$ represents the first $K$ columns of $U$, $\Delta$ and $V$, respectively:

$$\begin{aligned} U_{[k]} &= [u_1, \cdots, u_k, \cdots, u_K] \\ V_{[k]} &= [v_1, \cdots, v_k, \cdots, v_K] \\ \Delta_{[k]} &= [\delta_1, \cdots, \delta_k, \cdots, \delta_K] \end{aligned} \tag{22}$$

The hard-threshold (hard-THR) singular value process can be expressed as:

$$H_{[k]}^i = U_{[k]} \Delta_{[k]} V_{[k]}^T \tag{23}$$

where $H_{[k]}^i$ represents the matrix $H$ reconstructed from the first $K$ eigenvectors. The SVD is particularly useful for ill-posed linear problems with nearly degenerate matrices because it provides the best approximation with lower rank.

After low-rank processing, the Hankel matrix is transformed back to the sine image through the inverse Hankel transform (IHT) operation denoted by $H^+(\bullet)$, represented as follows:

$$x^i \leftarrow H^+(H_{[k]}^i) \tag{24}$$

***TV Step:*** TV minimization is also employed to remove noise and artifacts. Suppose $\Delta x = \left\| x - x^i \right\|$, TV minimization can be stated as follows:

$$TV(x^i) = x^{i+1} - \alpha \Delta x \frac{\nabla \| x^i \|_{TV}}{\| \nabla \| x^i \|_{TV} \|} \tag{25}$$

where $\alpha$ is the length of each gradient-descent step.

***DC Step:*** To improve the noise immunity, the statistical properties of the projection data can be incorporated into the objective function [29, 32]. Additionally, there exists a statistical method for sinogram denoising that utilizes the PWLS method to obtain the best assessment from noisy sine images. The PWLS prior is integrated into a regularized objective function that is expressed as:

$$x^i = \arg\min_x [\left\| y - x^{i+1} \right\|_W^2 + \lambda_1 \left\| x^{i+1} - H^+(H_{[K]}^{i+1}) \right\|_2^2 + \lambda_2 \left\| x^{i+1} \right\|_{TV}^2] \tag{26}$$

where hyperparameters $\lambda_1$ and $\lambda_2$ balance the trade-off among the terms of PWLS, LR and TV. $i = N-1, \cdots, 0$ denotes the iteration of outer loop. Specifically, the standard PWLS can be described as follows:

$$x^i = \arg\min_x [(x^{i+1} - y)^T W (x^{i+1} - y) + \mu R(x^{i+1})] \tag{27}$$

where superscript $T$ represents the transposing operation. Equation (25) can be further solved as:

$$x^i = \frac{W(y - x^{i+1}) + \mu R'(x^{i+1})}{W + \mu} \tag{28}$$

In order to decrease the influence of noise, the scale coefficient $\eta$ for system calibration is set to 22000.

$$W = diag\{w_i\} = diag\{1/\sigma_{x_i}^2\} = diag\left\{1/l_1 \exp(x_i/\eta)\right\} \tag{29}$$

Regarding the corrector, it transforms any initial sample $x(t)$ into the final sample $x(0)$ through the following process, which is referred to as Langevin dynamics:

$$\begin{aligned} x^{i,j} &\leftarrow x^{i,j-1} + \varepsilon_i s_\theta(x^{i,j-1}; \sigma_i) + \sqrt{2\varepsilon_i} z \\ i &= N-1, \cdots, 0, \quad j = 1, 2, \cdots, M \end{aligned} \tag{30}$$

where $\varepsilon_i > 0$ is the step size, and $z \sim \mathcal{N}(0,1)$ refers to a standard normal distribution. The above formulation is repeated for $i = N-1, \cdots, 0, j = 1, 2, \cdots, M$. The theory of Langevin dynamics guarantees that when $M \to \infty$ and $\varepsilon_i \to 0$, $x^i$ is a sample from $p_t(x)$ under designated conditions.

Sampling is not performed directly from the distribution $p(x)$, but rather through the posterior distribution $p(x \mid y)$ as explained in Section II. The DC operation can be viewed as a conditional term, which is incorporated into the sampling process of (28) and results in:

$$\begin{aligned} x^{i,j} &= x^{i,j-1} + \varepsilon_i \nabla_x [\log p_t(y \mid x^{i,j-1}) + \log p_t(x_{LR}^{i,j-1}) \\ &\quad + \log p_t(x_{TV}^{i,j-1})] + \sqrt{2\varepsilon_i} z \end{aligned} \tag{31}$$

Once the reconstructed projection $x$ is obtained, the final image $\tilde{I}$ is obtained:

$$\tilde{I} = T(x) \tag{32}$$

where $T(\bullet)$ stands for filtered back-projection (FBP).

**Algorithm 1** describes a detailed description of the training and iterative reconstruction algorithm for the PHD mod-el. The entire process of PHD model reconstruction consists of two loops. In the outer loop, a trained network is used for prediction, followed by correction in the inner loop. The predictor and corrector function together as a whole are utilized to generate the final samples. Additionally, after the operation of both the predictor and corrector, a data fidelity term is applied to ensure the quality of the generated images.

**Algorithm 1: iterative reconstruction of PHD**

**Training Stage**

**Dataset:** Few projection domain samples $x$

**1: Repeat**

**2:** $x \sim p(x)$, $t \sim \mathcal{U}([0,T])$, $\varepsilon \sim \mathcal{N}(0,I)$

**3:** $x(t) = x(0) + \varepsilon\sigma(t)$

**4:** Take a gradient descent step on $\nabla_\theta \left\| s_\theta(x(t),t) + \varepsilon \right\|_2^2$

**5: Until** converged

**6:** Trained PHD

**Iterative Reconstruction Stage**

**Setting:** $s_\theta, N, M, \sigma, \varepsilon$

**1:** Initial data $x = T^{-1}(I)$ **(FP)**

**2:** $x^N \sim \mathcal{N}\left(0, \sigma_{\max}^2 I\right)$

**3: For** $i = N-1$ to $0$ **do (Outer loop)**

**4:** $x^i \leftarrow Predictor(x^{i+1}, \sigma_i, \sigma_{i+1}, s_\theta)$

**5:** $[H_1^i, H_2^i] = H_p^i \leftarrow H(x^i)$ **(HT)**

**6:** $H_{vL}^i$, $H_{vR}^i$ via (18)

**7:** $H_3^i = [H_{1R}^i, H_{2L}^i]$

**8:** $\bar{H}_{1R}^i$, $\bar{H}_{2L}^i$ via (20)

**9:** $[U\Delta V^T] = svd[H_{1L}^i, \bar{H}_{1R}^i, \bar{H}_{2L}^i, H_{2L}^i]^T$ **(SVD)**

**10:** $H_{[k]}^i = U_{[k]}\Delta_{[k]}V_{[k]}^T$ **(hard-THR)**

**11:** $x^i \leftarrow H^+(H_{[k]}^i)$ **(IHT)**

**12:** $x^i = \dfrac{W(y - x^{i+1}) + \mu R'(x^{i+1})}{W + \mu}$ **(PWLS)**

**13:** $x^i = TV(x^i)$ **(TV)**

**14: For** $j = 1$ to $M$ **do (Inner loop)**

**15:** $x^{i,j} \leftarrow Corrector(x^{i,j-1}, \sigma_i, \varepsilon_i, s_\theta)$

**16: Repeat from step 5 to step 10**

**17: End for**

**18: End for**

**19:** Final image $\tilde{I} = T(x)$ **(FBP)**

**20: Return** $\tilde{I}$

The flowchart of the iterative reconstruction process is shown in Fig. 5. During the inference stage, iterations are performed between the iterative updates of the numerical SDE solver and the data consistency step to achieve reconstruction.

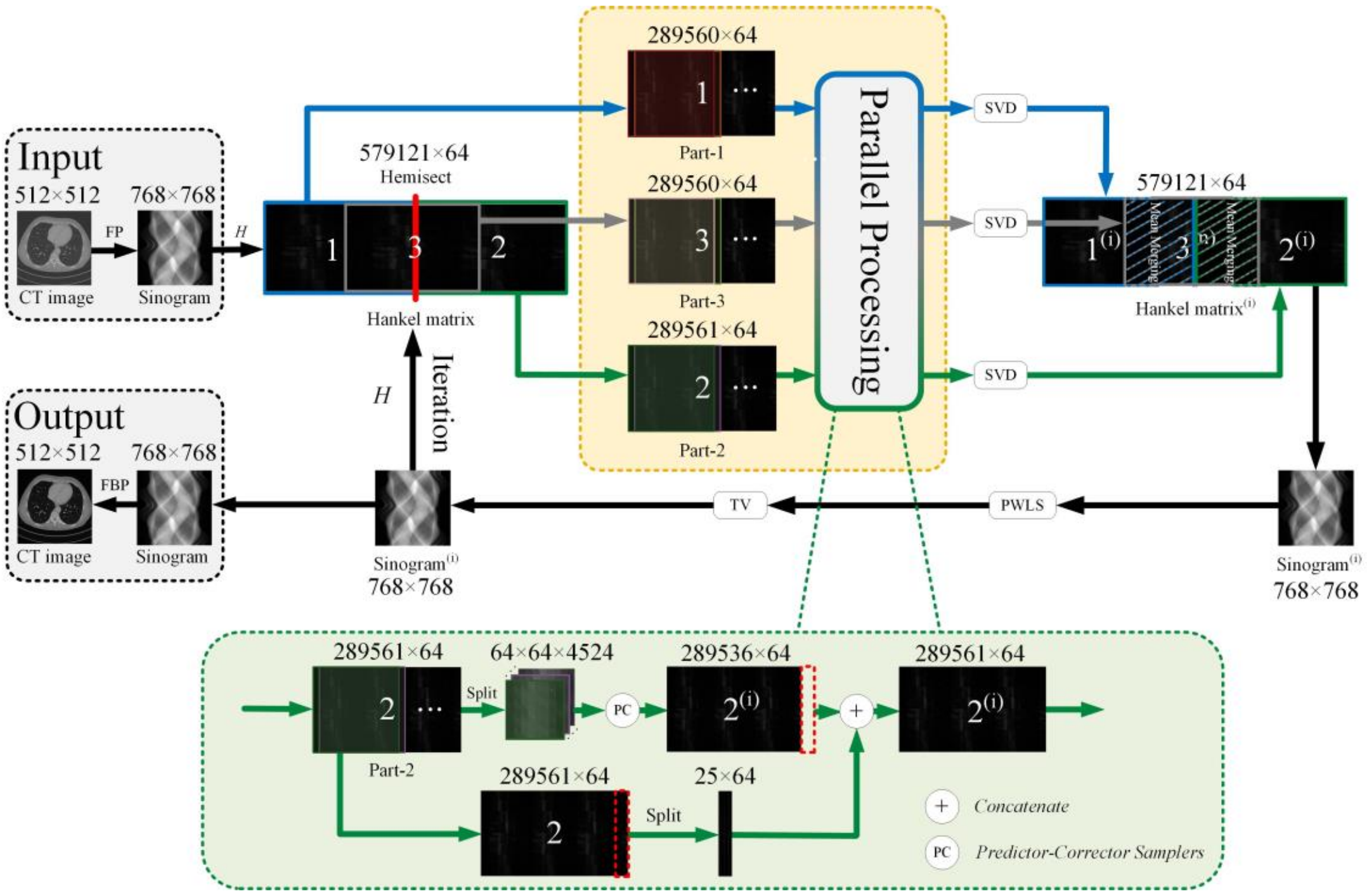

**Fig. 5.** The pipeline for iterative reconstruction stage of PHD. During the reconstruction stage, low-dose CT reconstruction is performed by employing parallel processing of multiple partitioned Hankel matrices.

## IV. Experiments

### A. Data Specification

***AAPM Challenge Dataset:*** Mayo Clinic provided simulated human abdominal images for the AAPM Low-Dose CT Grand Challenge to evaluate the performance of different algorithms in low-dose CT imaging [33]. The dataset consists of 2588 normal-dose CT images with a resolution of $512 \times 512$ and a slice thickness of 3mm from 10 patients. After processing, the images were transformed into 7764 normal-dose CT images with a resolution of $512 \times 512$ and a slice thickness of 1mm. Projection data with different noise levels (1e5, 5e4, and 1e4) were generated by adding Poisson noise to sinograms obtained from the normal-dose CT images. The artifact-free images generated from the normal-dose projection data using the FBP algorithm can be considered as ground truth. For fan-beam CT reconstruction, the Siddon's ray-driven algorithm [34] was utilized to generate the projection data. The distance from the rotation center to the source and detector was set to 40 cm. The detector width was 41.3 cm, composed of 720 detector elements. Over 360 projection views were uniformly distributed throughout the entire imaging process. In this study, a total of 4742 projection data were selected for the training set. All models were validated using 537 sinograms, and 12 sinograms were chosen as the test set.

***Somatom Confidence CT Dataset:*** The generalization study was evaluated on a subset of CT scans from the dataset proposed by Zeng *et al*. [35]. This dataset comprises data from 307 patients who underwent VMAT treatment, with approximately 100 radiation dose maps and 100 corresponding CT images per patient. The CT images were acquired using Somatom Confidence (Siemens healthcare, Forchheim, Germany) and were in DICOM format with a resolution of $256\times256\times160$ . The CT images were resampled to a resolution of $512\times512$, and 12 CT scan images were selected as the test set for the generalization study.

### B. *Model Training and Parameter Selection*

In the experiments, the PHD model is implemented in Python using Operator Discretization Library (ODL) [36] and PyTorch on a personal workstation with 2 GPUs (NVIDIA TITAN Xp-12GB). To accurately simulate real-world conditions, Poisson noise is added to the testing slices, with X-ray source intensities of $a_i = 1e5$, $a_i = 5e4$, and $a_i = 1e4$. During the training stage, the PHD model utilizes the Adam algorithm with a specific learning rate 0.001 and Kaiming initialization for weight initialization. The reconstruction process involves outer and inner iterations, with $N = 300$ outer iterations and $M = 2$ inner iterations. The correction process in the inner loop is iterated twice using annealing Langevin. $\eta$ in PWLS scheme is set to a constant 22000. The number of iterations $t$ in TV minimization is set to 2. The singular value thresholding in SVD operation is 38 and the sliding window size is $8\times8$ . For the convenience of reproducibility, the source code and some representative results are available at: *https://github.com/yqx7150/PHD*.

### C. *Evaluation Metrics*

The evaluation of reconstructed data quality involves the use of quantitative metrics such as peak signal-to-noise ratio (PSNR), structural similarity index (SSIM), and mean squared error (MSE). These metrics are calculated by averaging the results from 12 test data points.

PSNR describes the maximum possible power of the signal in relation to the noise corrupting power. Higher PSNR values correspond to better quality reconstructions. Denoting $I$ and $\tilde{I}$ to be the estimated reconstruction and ground-truth, PSNR is expressed as:

$$PSNR(I,\tilde{I}) = 20\log_{10}[MAX(\tilde{I})/\|I-\tilde{I}\|_2] \quad (33)$$

The SSIM value is used to measure the similarity between the ground-truth and reconstruction. SSIM is defined as:

$$SSIM(I,\tilde{I}) = \frac{(2\mu_I\mu_{\tilde{I}}+c_1)(2\sigma_{I\tilde{I}}+c_2)}{(\mu_I^2+\mu_{\tilde{I}}^2+c_1)(\sigma_I^2+\sigma_{\tilde{I}}^2+c_2)} \quad (34)$$

where $\mu_I$ and $\sigma_I^2$ are the average and variances of $I$ . $\sigma_{I\tilde{I}}$ is the covariance of $I$ and $\tilde{I}$ . $c_1$ and $c_2$ are used to maintain a stable constant. MSE is employed to evaluate the errors and defined as:

$$MSE(I,\tilde{I}) = \frac{1}{W}\sum_{i=1}^{W}\|I_i - \tilde{I}_i\|_2 \quad (35)$$

where $W$ is the number of pixels within the reconstruction result. If MSE approaches to zero, the reconstructed image is closer to the reference image.

### D. *Experimental Evaluation*

***AAPM Challenge Dataset Comparison:*** We compare the proposed unsupervised model PHD with five baseline techniques in low-dose CT reconstruction, including FBP [1], SART-TV [37], CNN [38], NCSN++ [16], and OSDM [17]. The CNN model is trained using 4742 paired projection data, similarly, the NCSN++ model is also trained using 4742 projection data. In addition, the OSDM and PHD model are trained with 1 and 50 projection data, respectively. The involved parameters are set following the guidelines in their original papers. This comparative analysis aims to assess the performance and effectiveness of these techniques in low-dose CT reconstruction.

In low-dose CT reconstruction, experiments with different noise levels are conducted by setting 1e5, 5e4, and 1e4 photons along each path of the X-ray. Table I shows the PSNR, SSIM, and MSE values of the reconstructed images. The best PSNR and MSE values are highlighted in bold. The reconstructed images produced by PHD exhibit fewer artifacts and less noise. It is evident that compared to FBP, PHD exhibits a significant increase in average PSNR in the noise levels of 1e5, 5e4, and 1e4, even when trained with only one sample. The gains are 8.67dB, 9.1dB, and 12.17dB, respectively.

TABLE I
RECONSTRUCTION PSNR/SSIM/MSE OF AAPM CHALLENGE DATA USING DIFFERENT METHODS AT DIFFERENT NOISE LEVEL.

| Noise level / Method (Samples) | $a_i$=1e5 | $a_i$=5e4 | $a_i$=1e4 |
|---|---|---|---|
| FBP | 34.62/0.9252/3.66e-4 | 32.43/0.8866/5.81e-4 | 25.78/0.6897/2.69e-3 |
| SART-TV | 41.03/0.9892/8.65e-5 | 38.72/0.9786/1.39e-4 | 29.58/0.8710/1.18e-3 |
| CNN | 41.35/0.9903/8.45e-5 | 39.24/0.9859/1.34e-4 | 37.55/**0.9793**/1.99e-4 |
| NCSN++ | 41.52/0.9873/7.19e-5 | 40.19/0.9810/9.66e-5 | 37.30/0.9637/1.88e-4 |
| OSDM | 42.62/0.9899/5.51e-5 | 41.20/0.9857/7.74e-5 | 37.43/0.9683/1.83e-4 |
| PHD (1) | 43.29/0.9906/4.71e-5 | 41.53/0.9861/7.10e-5 | 37.95/0.9711/1.62e-4 |
| PHD (50) | **43.34/0.9907/4.67e-5** | **41.60/0.9864/6.98e-5** | **38.01**/0.9716/**1.60e-4** |

In order to further highlight its advantages, Figs. 6-7 display reconstructed images and residual images under different levels of noise. In comparative methods, the FBP algorithm performs the worst due to its high sensitivity to noise, resulting in significant artifacts and incomplete structural organization in the reconstructed images. The performance of the SART-TV algorithm is better than FBP, but it still lacks complex structural details. Meanwhile, CNN sacrifices some edge details and overly smooths the edges. Compared to previous methods, both NCSN++ and OSDM exhibit improved reconstruction results, but they do not effectively capture texture details. In contrast, the images reconstructed by PHD exhibit preserved fine details and ensure the integrity of the structure.

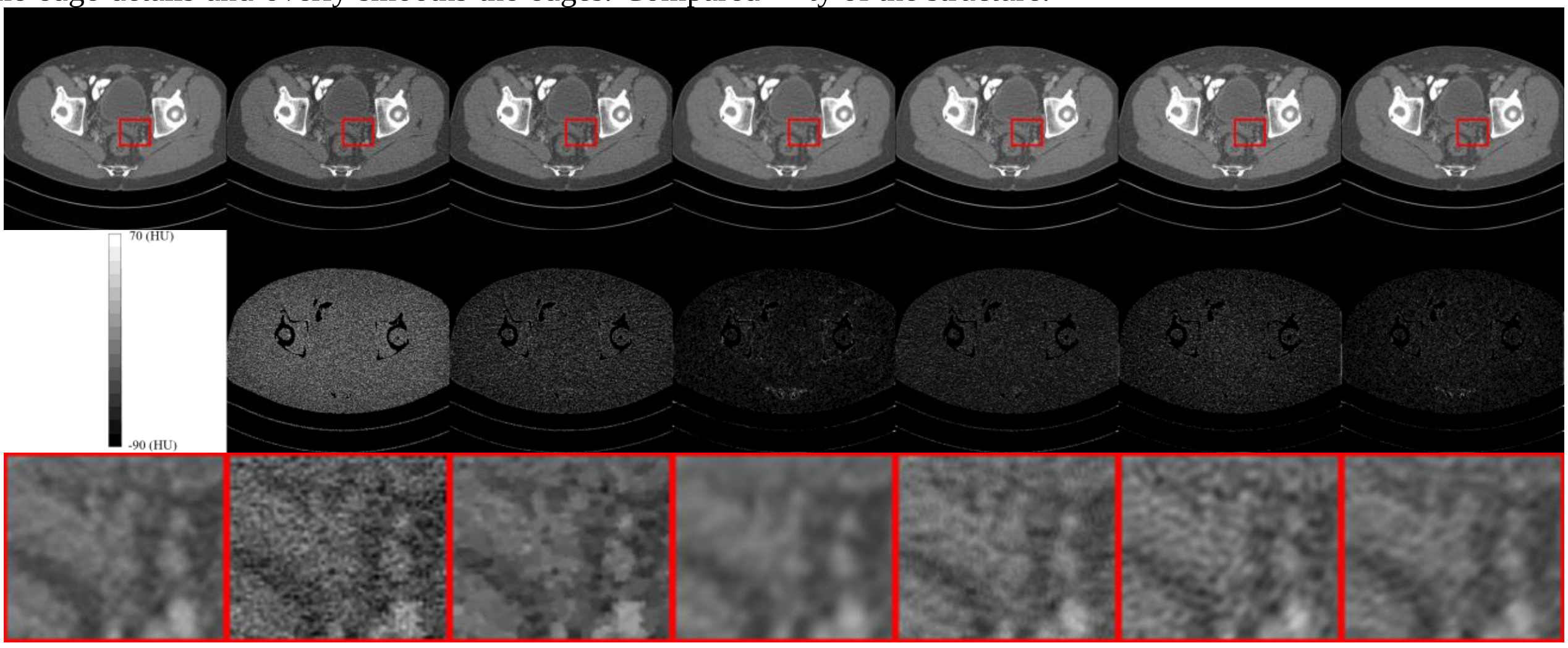

(a) (b) (c) (d) (e) (f) (g)

**Fig. 6.** Reconstruction results from 5e4 noise level using different methods. (a) The reference image versus the images reconstructed by (b) FBP, (c) SART-TV, (d) CNN, (e) NCSN++, (f) OSDM, (g) PHD (50). The display windows are [-90, 70] HU. The second row shows the error maps of the reconstruction, and the third row shows the enlarged view of the ROI (indicated by the red box in the first row).

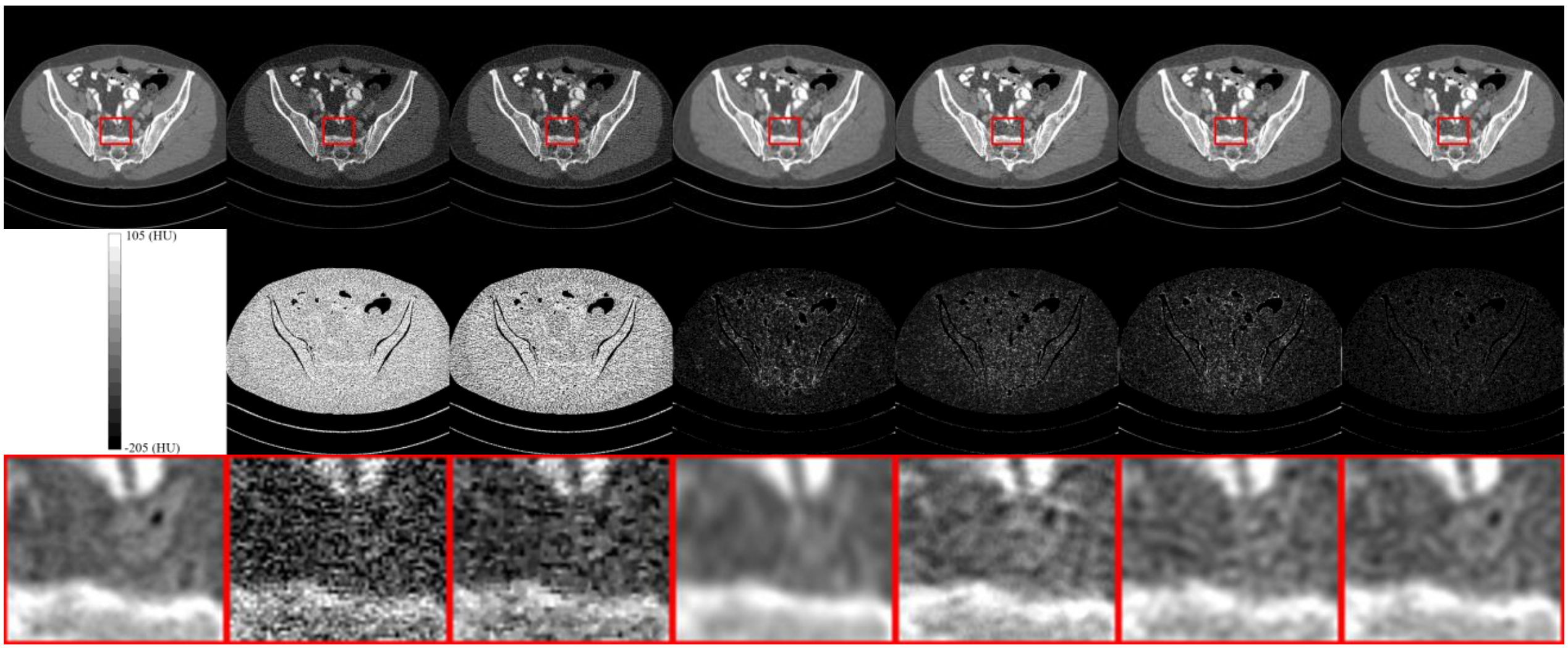

(a) (b) (c) (d) (e) (f) (g)

**Fig. 7.** Reconstruction results from 1e4 noise level using different methods. (a) The reference image versus the images reconstructed by (b) FBP, (c) SART-TV, (d) CNN, (e) NCSN++, (f) OSDM, (g) PHD (1). The display windows are [-205, 105] HU. The second row shows the error maps of the reconstruction, and the third row shows the enlarged view of the ROI (indicated by the red box in the first row).

After increasing the noise level from 5e4 to 1e4, there is a noticeable differentiation in the quality of the reconstructed images. Images generated by FBP and SART-TV exhibit prominent structural feature omissions and significant noise interference (Fig. 7(b) and Fig. 7(c)), while those generated by CNN tend to overly smooth the edges in the reconstructed images (Fig. 7(d)). Although the reconstruction quality of images by NCSN++ and OSDM improv, noticeable stripe artifacts still appear due to high noise interference (Fig. 7(e) and Fig. 7(f)). In contrast, under the high noise level of 1e4, PHD is able to preserve complex structural details to the maximum extent while effectively suppressing stripe artifacts (Fig. 7(g)). Furthermore, at this noise level, PHD demonstrates superior reconstruction performance compared to other contrast methods while being trained under the condition of maintaining a single sample, thus greatly reducing the demand for data samples. In summary, PHD proves highly effective in noise removal.

***Generalization Test:*** To better explore the generalization of the PHD model, we apply the comparative methods to the Somatom Confidence CT dataset. Furthermore, the pre-trained model with a noise level of 1e5 is used to reconstruct the CNN with noise levels of 1e4 and 5e2. The NCSN++, OSDM, and PHD still use the old models with prior knowledge obtained from the AAPM challenge dataset. The quantitative evaluations of PSNR, SSIM, and MSE metrics are presented in Table II. The PHD outperforms other methods in various indicators, demonstrating its superior performance. Additionally, Fig. 8 displays representative reconstruction results and error map of PHD and other methods. The PHD method achieves good reconstruction results and preserves distinct structural edge features.

TABLE II
RECONSTRUCTION PSNR/SSIM/MSE OF SOMATOM CONFIDENCE CT DATA USING DIFFERENT METHODS AT DIFFERENT NOISE LEVEL.

| Noise level | FBP | CNN | NCSN++ | OSDM | PHD (1) |
|---|---|---|---|---|---|
| $a_i$ =1e5 | 40.41/0.9763/9.31e-5 | 35.65/0.9773/4.45e-4 | 40.98/0.9855/8.91e-5 | 44.10/0.9911/4.09e-5 | **45.38/0.9938/3.05e-5** |
| $a_i$ =1e4 | 30.53/0.815/9.06e-4 | 34.63/0.9617/4.74e-4 | 35.86/0.9377/2.71e-4 | 38.09/0.9646/1.68e-4 | **40.02/0.9779/1.05e-4** |
| $a_i$ =5e2 | 19.42/0.2742/1.17e-2 | 26.86/0.6382/2.13e-3 | 28.06/0.7319/1.74e-3 | 30.67/0.7987/1.01e-3 | **32.88/0.8808/6.15e-4** |

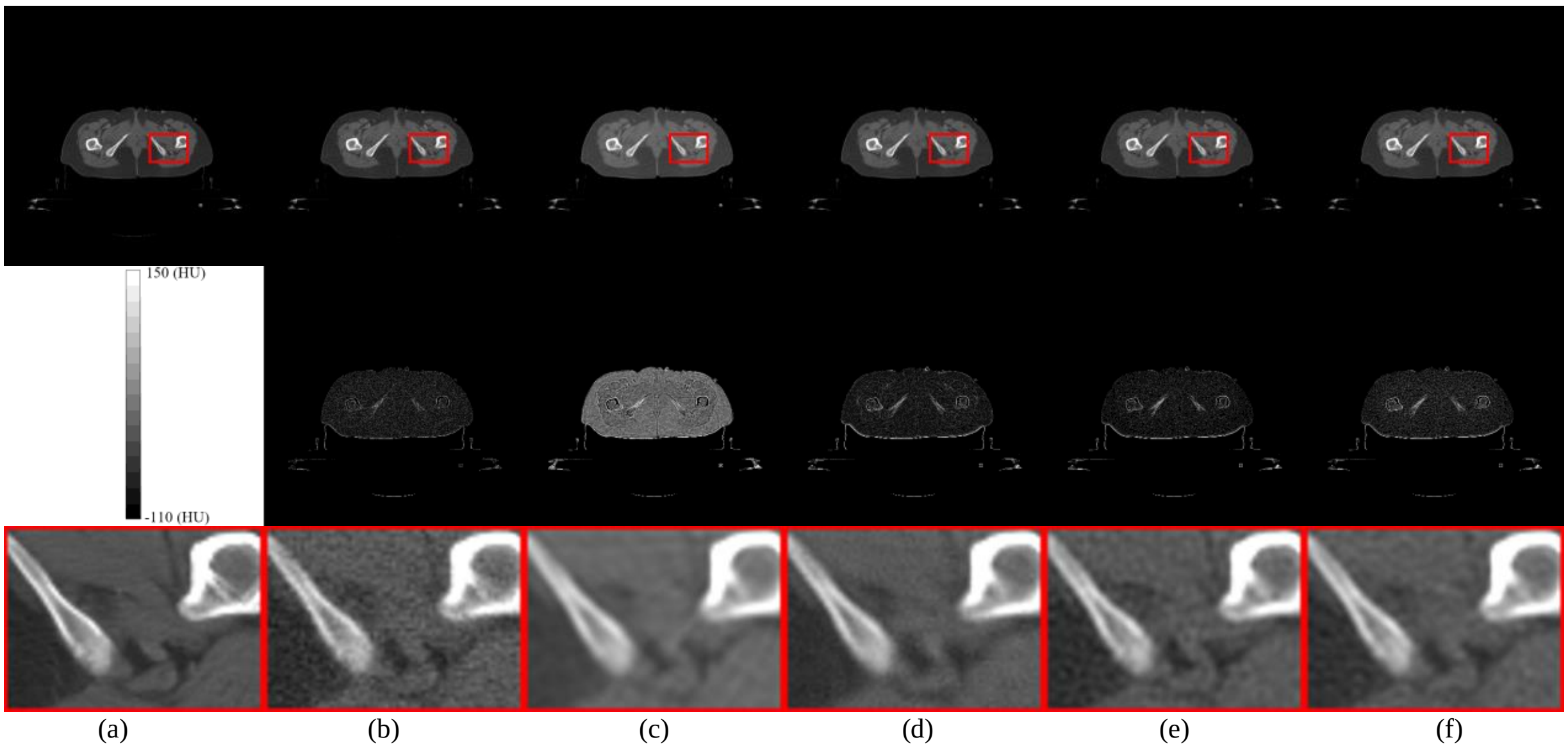

**Fig. 8.** Reconstruction results from 1e5 noise level using different methods. (a) The reference image versus the images reconstructed by (b) FBP, (c) CNN, (d) NCSN++, (e) OSDM, and (f) PHD (1). The display windows are [-110, 150] HU. The second row shows the error maps of the reconstruction, and the third row shows the enlarged view of the ROI (indicated by the red box in the first row).

### *E. Ablation Study*

We conduct ablation studies on multiple modules of the PHD and investigate the effectiveness of the proposed partitioned dimension in low-dose CT reconstruction.

***Exploring Partitioned Dimension:*** In this subsection, the impact of different partitioned dimension on the reconstruction performance is primarily investigated, as well as how to effectively improve within the same partitioned dimension. Furthermore, various metrics are employed in the experiments, including partitioned Hankel dimension (i.e., Size Per Hankel, SPH), in conjunction with PSNR, SSIM, and MSE, as shown in Table III.

TABLE III
COMPARISON OF PSNR/SSIM/MSE OF FOUR DIFFERENT PARTITIONED DIMENSIONS.

| Partitions | SPH | PSNR | SSIM | MSE |
|---|---|---|---|---|
| Dual | (289560,64) (289561,64) | 37.60 | 0.9694 | 1.79e-4 |
| Triple | (193040,64) (193040,64) (193041,64) | 37.71 | 0.9701 | 1.72e-4 |
| Triple* | (289560,64) (289560,64) (289561,64) | **37.95** | **0.9711** | **1.62e-4** |
| Quad | (144780,64) (144780,64) (144780,64) (144781,64) | 37.05 | 0.9665 | 2.03e-4 |

To ensure a fair comparison, in this experiment, the comparison conditions were standardized by training with the same single sample and testing with the same dataset. We perform low-dose CT reconstruction under a noise level of 1e4 and use dual-partition as the baseline. For the dual-partition (i.e., evenly splitting the Hankel matrix), the two sizes of the Hankel matrices are $289560\times64$ and $289561\times64$, respectively, and are reconstructed separately. Similarly, for the triple-partition, the Hankel matrix is divided into three equal parts, resulting in sizes of $193040\times64$, $193040\times64$, and $193041\times64$ for the Hankel matrices. When the triple-partition is extended to quad-partition, the Hankel matrix is divided into four equal parts, and the four resulting Hankel matrices are reconstructed separately.

The triple*-partition (PHD) further improves upon the triple-partition method by obtaining three Hankel matrices (part-1, part-2, part-3) with sizes $289560\times64$, $289561\times64$, and $289560\times64$, respectively. The specific configuration is illustrated in Fig. 5, where part-1 and part-2 are obtained from a Hankel matrix with a size of $579121\times64$. Additionally, a Hankel matrix with a size of $289560\times64$ is extracted from the middle and used as a new part (part-3). After separate reconstructions, part-3 is concatenated back to its original position. For the overlapping Hankel regions of part-1/2, a mean combination is performed to reconstruct a new Hankel matrix with a size of $579121\times64$.

As shown in Table III, the triple*-partition approach outperforms other partitioning schemes in terms of PSNR, SSIM and MSE, achieving optimal results. This is also the way adopted by the PHD.

The reconstruction results for different block partitioning dimensions are depicted in Fig. 9. It is evident that, under a high noise level of 1e4, the images generated by Dual/Triple/Quad-partition are all significantly affected by noise, resulting in poor preservation of texture details. Conversely, the triple*-partition demonstrates effective noise reduction, thereby minimizing artifacts and retaining structural details to the maximum extent.

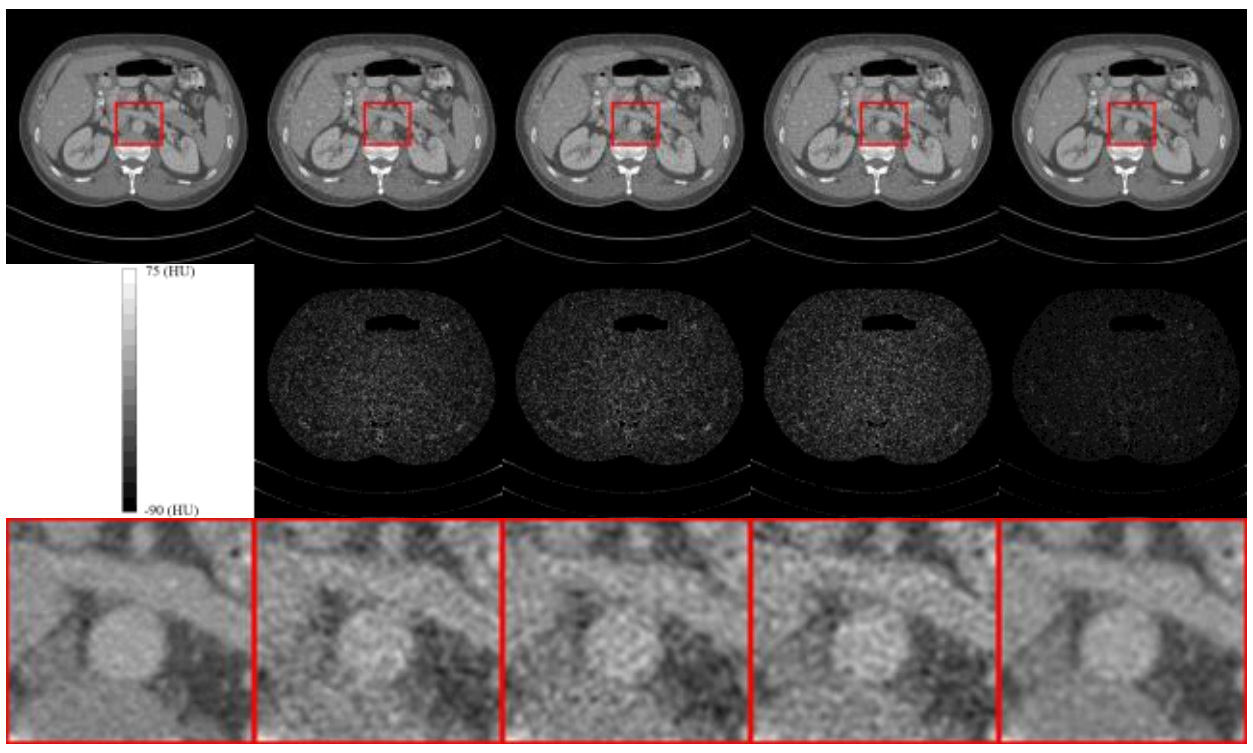

(a) (b) (c) (d) (e)

**Fig. 9.** Reconstruction results from 1e4 noise level using different partitioned dimensions. (a) The reference image versus the images reconstructed by (b) Dual-partition, (c) Triple-partition, (d) Quad-partition, (e) Triple*-partition. The display windows are [-90, 75] HU. The second row shows the error maps of the reconstruction, and the third row shows the enlarged view of the ROI (indicated by the red box in the first row).

***Different Components in PHD Model:*** To validate the significance of the corresponding modules in the PHD model, this study qualitatively and quantitatively compares it with the Hankel-k-space generative model (HKGM) [28] and HKGM with PWLS. All methods employ Hankel matrices to enhance data redundancy and mitigate the issue of limited data samples. Additionally, the PHD model incorporates a TV minimization module to improve image denoising and reconstruction.

The quantitative results for the reconstructions at noise levels of 1e5, 5e4, and 1e4 are displayed in Table IV. It is evident that the PHD model yields the best results in terms of PSNR, SSIM and MSE values, which are consistent with the visual effects.

TABLE IV
RECONSTRUCTION PSNR/SSIM/MSE OF AAPM CHALLENGE DATA USING DIFFERENT METHODS AT DIFFERENT NOISE LEVEL.

| Noise level | HKGM (50) | HKGM with PWLS (50) | PHD (50) |
|---|---|---|---|
| $a_i$ =1e5 | 42.28±0.49/0.9889±0.0015/5.95e-5 | 42.69±0.55/0.9892±0.0012/5.43e-5 | **43.34±0.49/0.9907±0.0008/4.67e-5** |
| $a_i$ =5e4 | 40.71±0.49/0.9841±0.0015/8.56e-5 | 40.89±0.60/0.9846±0.0014/8.21e-5 | **41.60±0.60/0.9864±0.0019/6.98e-5** |
| $a_i$ =1e4 | 35.64±0.67/0.9515±0.0066/2.76e-4 | 36.16±0.57/0.9541±0.0053/2.44e-4 | **38.01±0.65/0.9716±0.0035/1.60e-4** |

Fig. 10 illustrates the reconstructed images obtained using different methods at a noise level of 1e4. The reconstructed image generated by HKGM is heavily influenced by noise, resulting in the degradation of structural details and a significant presence of artifacts. However, although the HKGM model incorporates the PWLS module, which improves the edge texture, the denoising effect is still unsatisfactory. In contrast, PHD demonstrates excellent denoising capabilities while better preserving the characteristics of texture details. Based on conducted ablation studies, it is found that modules such as HKGM, PWLS, and TV minimization play crucial roles in the PHD model.

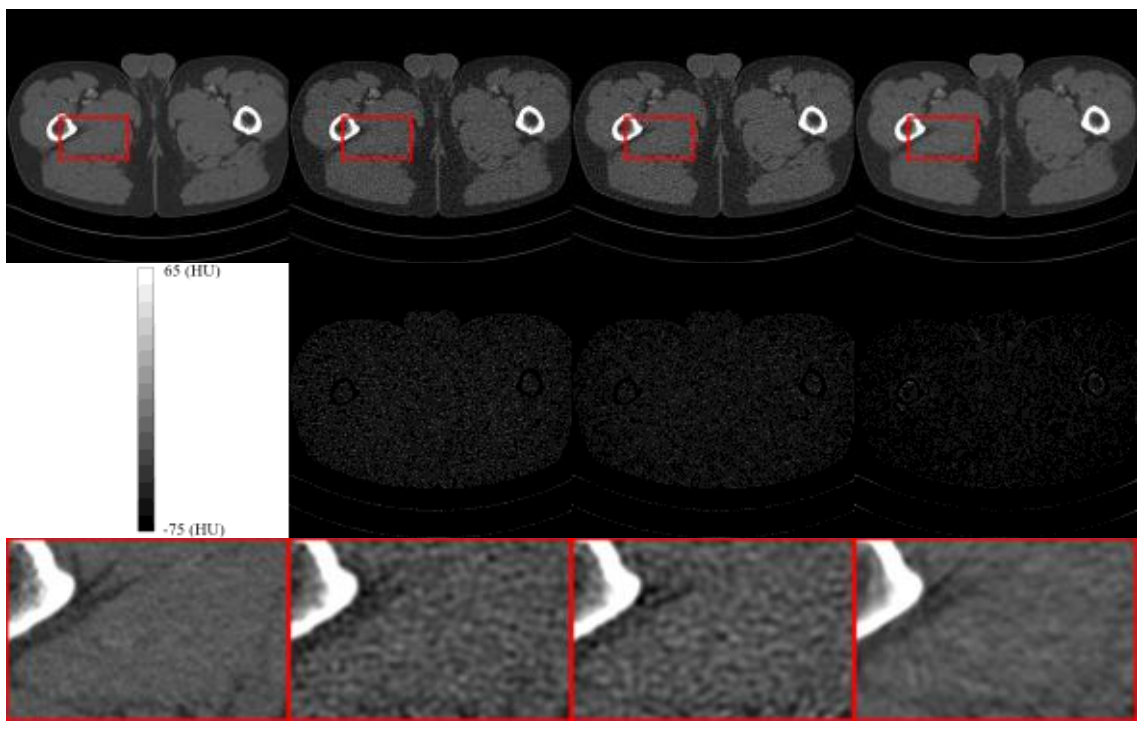

(a) (b) (c) (d)

**Fig. 10.** Reconstruction results from 1e4 noise level using different methods. (a) The reference image versus the images reconstructed by (b) HKGM, (c) HKGM with PWLS, (d) PHD (50). The display windows are [-75, 65] HU. The second row shows the error maps of the reconstruction, and the third row shows the enlarged view of the ROI (indicated by the red box in the first row).

## V. DISCUSSION

The input of the PHD model shifts from pure noise to low-dose CT projection data, significantly reducing the number of iterations with the assistance of the TV denoising module, as shown in Fig. 11. Significantly, the PSNR of the PHD model reaches its peak at the 10th step, whereas most diffusion models typically require 500 to 2000 steps [16-17, 39-40]. To achieve faster convergence and minimize the required time, we enforce TV denoising operations during the iterative reconstruction process of the PHD model. While a common practice for many image reconstruction diffusion models involves inputting under sampled images alongside introducing Gaussian noise to aid the reconstruction process, our observations during testing reveal that providing solely low-dose CT projection data without additional Gaussian noise enables the network to reach the desired outcome more rapidly. This approach not only accelerates the entire process but also notably reduces the required number of iterations.

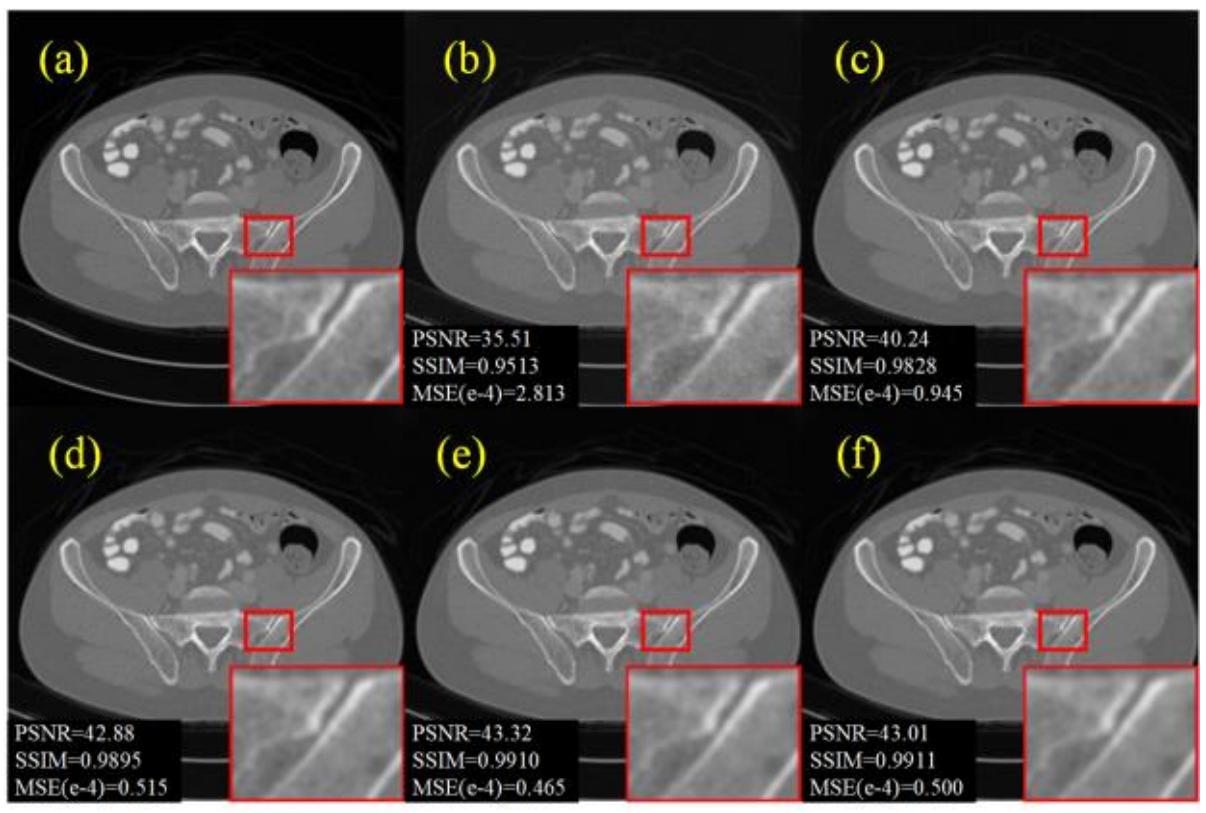

**Fig. 11.** Reconstruction steps for PHD. (a) Ground truth, (b) LDCT image, (c) Step=1, (d) Step=5, (e) Step=10, (f) Step=15. Quantitative results are provided in the bottom-left corner. The tissue boundaries of the images are enlarged in the bottom-right corner.

## VI. CONCLUSIONS

In recent years, deep learning-based CT reconstruction methods had made rapid progress. However, challenges still existed in terms of the generalizability and robustness of training networks. In this study, we introduced the Partitioned Hankel-based Diffusion models for low-dose CT re-construction using PC sampling to generate sinogram. This method employed an unsupervised approach to train the fraction-based generative model, better capturing the prior distribution of sinogram data. In the iterative reconstruction stage, numerical SDE solver, PWLS fidelity term and TV regularization steps were alternately performed. Additionally, a triple-partition parallel processing approach was adopted during both training and testing phases. The generalization and effectiveness of PHD were validated using the AAPM Challenge dataset and Somatom Confidence CT dataset. Moreover, considering the challenges of data collection in the medical field, the scarcity of data samples may reduce the effectiveness of model reconstruction. In future work, we plan to combine multi-domain and multi-model methods to fully utilize prior knowledge from a small number of samples and further improve the accuracy of reconstructed image.

## ACKNOWLEDGMENTS

The authors sincerely thank the anonymous referees for their valuable comments on this work. All authors declare that they have no known conflicts of interest in terms of competing financial interests or personal relationships.